\documentclass[aps,preprint,superscriptaddress,showpacs,amsmath,amssymb]{revtex4}

\usepackage{graphicx}
\usepackage{dcolumn}
\usepackage{bm}
\usepackage{color}
\usepackage{tabularx}
\usepackage{csquotes}
\usepackage[normalem]{ulem}

\newcommand{\cri}{CrI$_3$}

\newcommand{\tc}{$T_{\rm C}$}
\newcommand{\ts}{$T_{\rm S}$}
\newcommand{\tn}{$T^*$}
\newcommand{\aab}{$\alpha_{\rm ab}$}
\newcommand{\ac}{$\alpha_{\rm c}$}

\newcommand{\etal}{{\it et al.}}

\newcommand{\beginsupplement}{%
        \setcounter{table}{0}
        \renewcommand{\thetable}{S\arabic{table}}%
        \setcounter{figure}{0}
        \renewcommand{\thefigure}{S\arabic{figure}}%
     }

\begin{document}


\title{Uniaxial pressure effects in the two-dimensional van-der-Waals ferromagnet CrI$_3$}

\author{J.~Arneth}
\affiliation{Kirchhoff Institute of Physics, Heidelberg University, INF 227, D-69120 Heidelberg, Germany}
\author{M.~Jonak}
\affiliation{Kirchhoff Institute of Physics, Heidelberg University, INF 227, D-69120 Heidelberg, Germany}
\author{S.~Spachmann}
\affiliation{Kirchhoff Institute of Physics, Heidelberg University, INF 227, D-69120 Heidelberg, Germany}
\author{M.~Abdel-Hafiez}
\affiliation{Department of Physics and Astronomy, Uppsala University, Box 516, SE-751 20 Uppsala, Sweden}
\author{Y.O.~Kvashnin}\email{yaroslav.kvashnin@physics.uu.se}
\affiliation{Uppsala University, Department of Physics and Astronomy, Division of Materials Theory, Box 516, SE-751 20 Uppsala, Sweden}
\author{R.~Klingeler}\email{klingeler@kip.uni-heidelberg.de}
\affiliation{Kirchhoff Institute of Physics, Heidelberg University, INF 227, D-69120 Heidelberg, Germany}

\date{\today}

\begin{abstract}
Magnetoelastic coupling and uniaxial pressure dependencies of the ferromagnetic ordering temperature in the quasi-two-dimensional layered van-der-Waals material CrI$_3$ are experimentally studied and quantified by high-resolution dilatometry. Clear anomalies in the thermal expansion coefficients at $T_{\rm C}$ imply positive (negative) pressure dependencies $\partial T_{\rm C}/\partial p_{\rm i}$ for pressure applied along (perpendicular to) the $c$ axis. The experimental results are backed up by numerical studies showing that the dominant, intra-layer magnetic coupling increases upon compression along the $c$ direction and decreases with negative in-plane strain. In contrast, inter-layer exchange is shown to initially increase and subsequently decrease upon the application of both out-of-plane and in-plane compression.

\end{abstract}
\maketitle

\section{Introduction}
Quasi-two-dimensional (quasi-2D) layered van-der-Waals (vdW) materials have been intensively studied in the last years due to their rich physics including long-range-ordered phases down to the single layer.~\cite{huang2017, gong2017, OHara2018, sun2020} Thanks to the layered structure, they allow to address fundamentals of low-dimensional physics and also hold an outstanding promise for technological applications, as demonstrated, e.g., by Cr$_2$Ge$_2$O$_6$/NiO heterostructures or NiPS$_3$-based field-effect transistors.~\cite{jenjeti2018, idzuchi2019} In \cri , ferromagnetism emerges in the bulk at \tc\ = 61~K, and persists even down to the monolayer thickness with a slightly reduced ordering temperature.~\cite{huang2017} As magnetic long-range order in isotropic 2D spin systems is supposed to be suppressed by thermal fluctuations according to the Mermin-Wagner theorem~\cite{mervinwagner}, the presence of intrinsic ferromagnetism below about 45~K in single layers of \cri\ raises the question of its driving origin(s). 

The coupling between localized spins in CrI$_3$ is governed by different physical processes. 
The nature of isotropic (Heisenberg) interactions, which are usually dominant in materials with localized spins, is relatively well understood.~\cite{besbes2019,soriano2020}
At the same time, the role of spin-orbit coupling is not completely elucidated.
It is clearly important, as it provides a necessary source of magnetic anisotropy which breaks the spin rotational invariance, opening the gap in the magnon excitations and thus allowing for the long-range magnetic order to exist at finite temperatures.~\cite{chen2020} 
However, the exact shape of the spin Hamiltonian describing bulk CrI$_3$ is debated. Different studies suggest the importance of Dzyaloshinskii-Moriya~\cite{chen2018}, Kitaev~\cite{xu2018,lee2020}, or even non-relativistic higher-order interactions.~\cite{kartsev2020}
As previously discussed\cite{chen2020}, several models are equally successful in describing the inelastic neutron scattering data including the gap of $\Delta \approx 5$~meV at the Dirac points, which points towards the existence of highly intriguing topological magnons in this material. Further experiments are needed to fully understand the nature of ferromagnetic order in CrI$_3$ and exploit it in applications.
Magnetism in CrI$_3$ is also closely related to the crystal structure. At \ts\ $\approx 220$~K, \cri\ features a discontinuous phase transition from the high-temperature monoclinic (C2/m) to the low-temperature rhombohedral (R$\overline{3}$) phase.~\cite{mcguire2015} This transition is absent in few-layer systems and different stacking patterns appear in thin and bulk samples at low temperatures, thereby affecting the evolution of magnetic order.~\cite{ubrig2019, sivadas2018}

Here, we report high-resolution thermal expansion on bulk single crystals of CrI$_3$. The data show clear anomalies at the onset of bulk ferromagnetism which allow the quantitative determination of uniaxial pressure dependencies of \tc . Furthermore, we compare our experimental results with numerical calculations on the strain dependence of the dominant magnetic exchange couplings to elucidate a microscopic picture of the mechanisms governing ferromagnetism in bulk \cri . 

\section{Methods}
The experiments were performed on \cri\ single crystals from HQ Graphene~\footnote{www.hqgraphene.com} which display a structural transition at \ts\ $\simeq 212$~K and  ferromagnetic order at \tc\ = 61~K. The magnetization was studied by means of a Physical Properties Measurement System (PPMS, Quantum Design) using the vibrating sample magnetometer (VSM) option. High-resolution dilatometry measurements were performed by means of a three-terminal high-resolution capacitance dilatometer in a home-built setup placed inside a Variable Temperature Insert (VTI) of an Oxford magnet system.~\cite{kuechler2012,werner2017} With this dilatometer, the relative length changes $dL_i/L_i$ out-of-plane and in-plane, i.e., along the crystallographic $c$-direction and in the $ab$-plane, respectively, were studied on thin single crystals of dimensions $1.2 \times 1.8 \times 0.06~$mm$^{3}$. The thermal expansion measurements were performed upon warming at a rate of $0.3~$K/min.
A point-by-point derivative of the pre-processed data yields the linear thermal expansion coefficients $\alpha_i=1/L_i\cdot dL_i(T)/dT$ ($i = c, \perp c$). Due to air sensitivity of the crystals, fresh samples from the same batch were utilized for the respective measurements.

For numerical studies of strain effects, the equilibrium crystal structure of bulk CrI$_3$ was obtained by performing a complete structural optimization within density functional theory (DFT).
As for the initial guess, we adopted the parameters of the experimental low-temperature structure and kept the same point group symmetry (R$\bar{3}$).~\cite{mcguire2015}
The lattice parameters and atomic positions were relaxed using the PBE\cite{gga-pbe} functional by means of a projector augmented wave method as implemented in the VASP code.\cite{paw,vasp}
The plane-wave kinetic energy cut-off was set to 350~eV along with a 12$\times$12$\times$12 $k$-point grid.
The forces on each atom were minimized down to 1~meV/\AA.
Once the equilibrium structure was obtained, we applied the strain within the plane of the magnetic layers ($ab$) as well as along the perpendicular direction ($c$ axis). 
For every chosen value of strain, the positions of all the atoms in the material were optimized.

In order to extract magnetic interactions, the system was mapped onto a Heisenberg model: 
\begin{eqnarray}
\hat H  = -\sum_{\langle i,j\rangle} J_{ij} {\bf e}_i \cdot {\bf e}_j,
\label{heis-ham}
\end{eqnarray}
where ${\bf e}_i$ is the unit vector pointing along the spin moment of a Cr$^{3+}$ ion and $J_{ij}$ is the exchange interaction between Cr atoms at lattice sites $i$ and $j$.
In this sum, we restrict ourselves to the nearest-neighbour (NN) coupling $J_1$ and inter-layer coupling $J_\perp$.
In order to calculate the exchange coupling between two selected atoms, we calculated the energies of four magnetic configurations: 
$\mid \cdot \cdot \uparrow \cdot \cdot \uparrow \rangle $,  
$\mid \cdot \cdot \uparrow \cdot \cdot \downarrow \rangle$,  
$\mid \cdot \cdot \downarrow \cdot \cdot \uparrow \rangle$,  
$\mid \cdot \cdot \downarrow \cdot \cdot \downarrow \rangle$, which we denote as 
$E^{\uparrow\uparrow}$,
$E^{\uparrow\downarrow}$, 
$E^{\downarrow\uparrow}$, 
$E^{\downarrow\downarrow}$, respectively.
The rest of the spins were pointing "up", thus representing the FM background, corresponding to the ground state in bulk \cri .
This way, the exchange coupling was calculated as follows:
\begin{eqnarray}
J_{ij}=\frac{1}{8z} \biggl[ E^{\uparrow\downarrow} + E^{\downarrow\uparrow} - (E^{\uparrow\uparrow} + E^{\downarrow\downarrow}) \biggl],
\end{eqnarray}
where $z$ is the number of equivalent neighbours $j$ for a given site $i$.
The values of $J_1$ and $J_\perp$ were obtained for several values of in-plane and out-of-plane strain.
In order to minimize the effect of more distant couplings, these calculations were performed in an orthorhombic $(\sqrt{3}a,~a,~c)$ supercell containing 48 atoms.

There is a number of previous studies which used the same methodology to assess the magnetic exchange in Cr$X_3$ systems both in bulk and monolayer forms.~\cite{lu2019, lado2017, webster2018, xu2020, xu2018, olsen2019, xue2019}
The effect of strain in this class of materials was studied, for instance, in Refs.~\onlinecite{zhang2015, webster2018, pizzochero2020}.

\section{Results and discussion}

\begin{figure}[ht]
\includegraphics[width=1\columnwidth,clip]{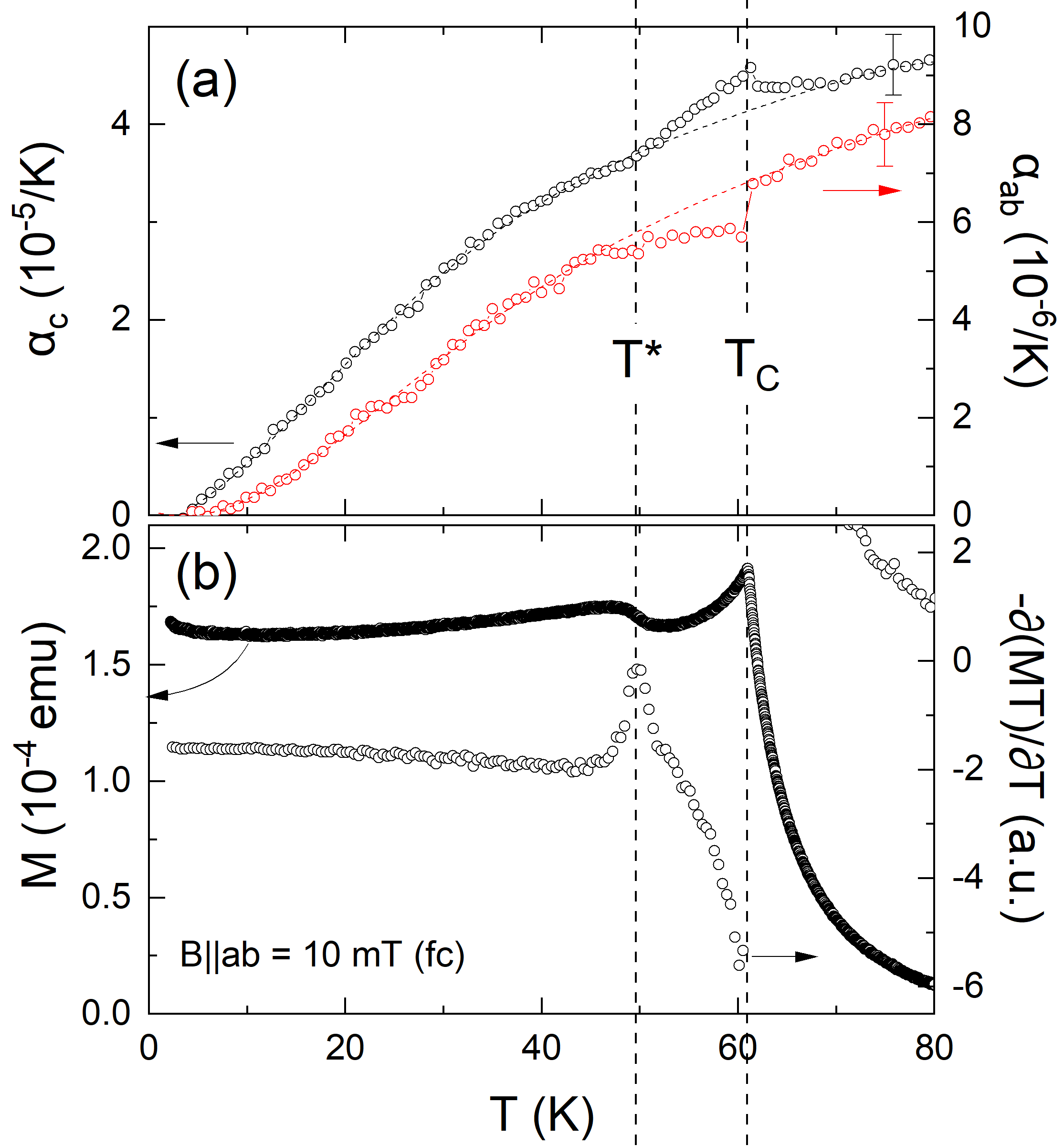}
\caption{(a) Thermal expansion coefficients of \cri , as well as (b) magnetization and its derivative, $-\partial (M\cdot T)/\partial T$, obtained at $B\parallel ab=10$~mT in a field-cooled measurement. Dashed lines mark the ferromagnetic ordering temperature \tc\ and the evolution of a surface antiferromagnetic/bulk ferromagnetic phase at \tn .}\label{lt}
\end{figure}

The thermal expansion coefficients presented in Fig.~\ref{lt}a display anomalies both in \ac\ and \aab\ at the ferromagnetic ordering temperature, thereby evidencing magnetoelastic coupling in \cri . While the anomaly temperatures coincide with the peak in magnetization at \tc , the anomaly in the magnetization at $T^*=49.5$~K signaling the evolution of antiferromagnetic ordering of the surface layers~\cite{niu2020, li2020, mccreary2020, liu2019} is not associated with clear anomalies in the thermal expansion. In addition, the out-of-plane thermal expansion also displays an anomaly at the structural transition temperature \ts\ $\simeq 215$~K. However, while the data are well reproducible below 150~K, this is not the case around \ts . We attribute this to irreversible stacking defects appearing in the sample at \ts\ and, hence, restrict the discussion to the temperature regime below 150~K.~\cite{mcguire2015} In contrast, the in-plane thermal expansion data are fully reproducible and do not show any detectable anomaly in \aab\ at \ts , i.e., there are no anomaly-related in-plane lattice changes in the detection limit $dL \approx 10^{-10}\,\mathrm{m}$ (see Supplement Fig.~S1).~\cite{mcguire2015} Comparing in-plane and out-of-plane data, thermal expansion in \cri{} is considerably larger along the $c$-axis as compared to the $ab$-plane, which is associated with very distinct directional chemical bonding. A comparably strong anisotropic behavior of the thermal expansion coefficients has been observed, for instance, in graphite, although \aab\ becomes negative at low temperatures.~\cite{janete2021}

\begin{figure}[thb]
\includegraphics[width=1\columnwidth,clip]{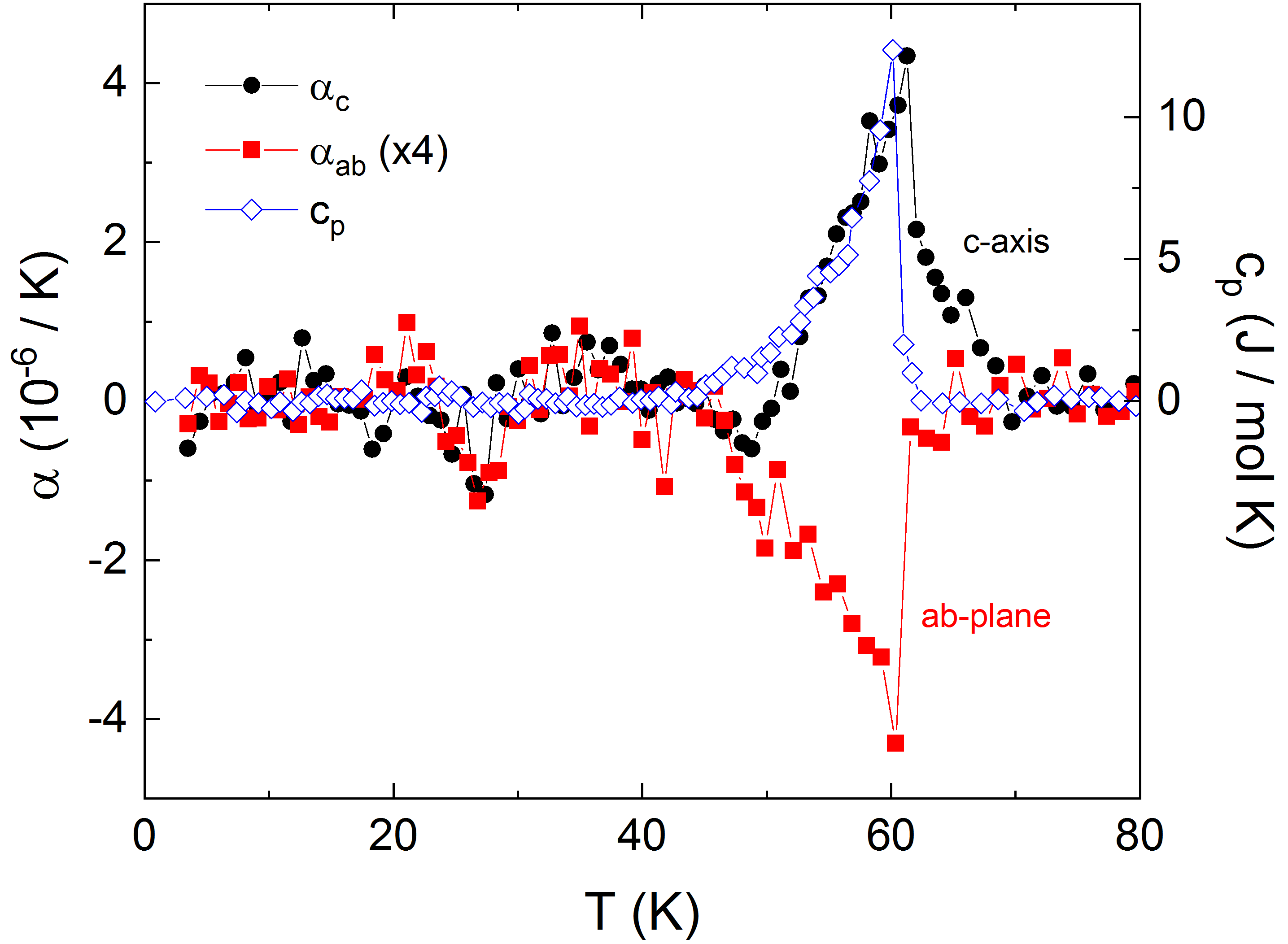}
\caption{Anomalous contributions to the thermal expansion coefficients and the specific heat from Ref.~\onlinecite{spurgeon2020} (see the text). The data \aab\ are multiplied by a factor 4.}\label{ano}
\end{figure}

At \tc\ = 61.5(5)~K, the small but distinct thermal expansion anomalies in \ac\ and \aab\ exhibit opposite signs, which implies opposite uniaxial pressure dependencies of \tc , i.e., $\partial T_{\rm C}/\partial p_{\rm c}>0$ and $\partial T_{\rm C}/\partial p_{\rm ab}<0$. To estimate the size of the anomalies, we fitted the thermal expansion coefficients well above \tc\ and well below \tn\ by a polynomial which is indicated in Fig.~\ref{lt}a. Anomalous contributions to the thermal expansion coefficients are obtained by subtracting the polynomial background from the experimental data (see Fig.~\ref{ano}). Note that, by construction, this phenomenological background includes putative length effects of magnetic short-range correlations which have been observed above \tc .~\cite{mcguire2015,spurgeon2020}

The anomaly at \tc\ is about four times larger in \ac\ than in \aab, which implies that the corresponding in-plane and out-of-plane pressure derivatives of \tc\ differ by that factor, as well. Applying the same procedure for background determination to the specific heat data from Spurgeon \etal~\cite{spurgeon2020} reveals a quite similar temperature dependence of the anomalous contributions to $\alpha_i$ and to $c_p$ (see Fig.~\ref{ano}), thereby signaling a temperature-independent Grüneisen ratio $\gamma_i = \alpha_i/c_p$, at the ferromagnetic phase transition. This experimental observation implies the presence of a single dominant energy scale $\epsilon^*$ which enables us to exploit the Grüneisen relation $V_{\rm m}\gamma_i = \partial \ln\epsilon^* /\partial p_i$.~\cite{lorenz2007,heyer2011,klingeler2006} Moreover, identifying $\epsilon^*$ with the ordering temperature $T_\mathrm{C}$ provides the uniaxial pressure dependencies of $T_\mathrm{C}$ which are listed in table~\ref{tab}\footnote{Error bars in the table stem from the macroscopic sample length, background determination, and scaling.}.~\cite{gegenwart2016,klingeler2006,heyer2011,hoffmann2021}

In an alternative approach to determine $\partial T_{\rm C}/\partial p_i$ from the data, we have deduced the actual jumps in the thermal expansion coefficients $\Delta \alpha_{\rm i}$ and in the specific heat $\Delta c_{\rm p}$ at the ferromagnetic transition by means of area-conserving constructions and used the Ehrenfest relation $dT_{\rm C}/dp_i = T_{\rm C} V_{\rm m} \Delta\alpha_{\rm i} /\Delta c_{\rm p}$.~\cite{kuechlerDiss,meingast1990,majumder2018} As shown in Table~\ref{tab}, this procedure yields similar results as the above-mentioned Gr{\"u}neisen analysis. 

\begin{table}[b]
    \centering
    \begin{tabular}{c||c|c|c|c}
            &$\gamma_\mathrm{i}$ & $\partial \ln \epsilon^*  /\partial p_i$ & $\partial T_{\rm C}/\partial p_i$  & $T_{\rm C} V_{\rm m} \Delta\alpha_\mathrm{i} /\Delta c_{\rm p}$   \\ 
            & (Mol/J) & ($10^{-2}$/GPa) & (K/GPa) & (K/GPa)   \\  \hline
            $p~\|~c$ & $3.5\times 10^{-7}$ &  2.8(5) & 1.7(3) & 1.5(6) \\
            $p\perp c$ & $-8.7\times 10^{-8}$ & 0.7(2) & -0.4(1) & -0.4(1) \\
    \end{tabular}
    \caption{Gr{\"u}neisen parameters and uniaxial pressure dependencies of $\ln \epsilon^*$ and \tc\ as obtained from the Gr{\"u}neisen parameter (middle columns) and from the anomalies $\Delta \alpha_i$ and $\Delta c_{\rm p}$ (right column), for pressure applied along and perpendicularly to the $c$-axis  (see the text).}
    

    \label{tab}
\end{table}

While uniaxial pressure dependencies for \cri\ have not been reported yet, two recent studies confirm monotonic increase of \tc\ upon the application of hydrostatic pressure.~\cite{ghosh2021arxiv,mondal2019} Note that the thermodynamic analyses presented above yield the respective initial pressure dependencies, i.e., for $p_i\rightarrow 0$. The initial pressure coefficient $\partial T_\mathrm{C}/\partial p \sim 1.3$~GPa reported by Ghosh \etal ~\cite{ghosh2021arxiv} is in good agreement with the results of our analysis in which we find
$\partial T_\mathrm{C}/\partial p = \partial T_\mathrm{C}/\partial p_c + 2 \times \partial T_\mathrm{C}/\partial p_{ab}\approx 0.9$~K/GPa.~\cite{he2018} In contrast, Mondal \etal\ report a value that is roughly six times larger.~\cite{mondal2019} However, direct comparison between the presently reported pressure dependence obtained from experimental $uniaxial$ measurements with \textit{hydrostatic} pressure experiments from the literature is not straightforwardly possible. Due to the strong structural anisotropy, the effect of hydrostatic pressure is not \textit{a priori} clear and supposed to mainly modify the inter-layer structure, such as the layer separation and stacking order, thereby effectively reflecting mainly uniaxial pressure along the $c$ axis.~\cite{li2019} 
In general, pressure effects reported for quasi-2D honeycomb magnets in the literature vary by almost three orders of magnitude. In $\alpha$-RuCl$_3$, hydrostatic pressure initially suppresses long-range antiferromagnetic order at a rate of $\sim 22$~K/GPa.~\cite{he2018} Whereas, the ordering temperature in the $S = 1$ Heisenberg system Na$_3$Ni$_2$SbO$_6$ is suppressed by only $\sim 0.05$~K/GPa.~\cite{werner2017} The pressure coefficient obtained from the analysis of our data on \cri\ is in between these values and is comparable to $\partial T_\mathrm{N}/\partial p \sim 0.7$~K/GPa found by dilatometric studies on the Kitaev iridate $\beta$-Li$_2$IrO$_3$.~\cite{majumder2018}


In the following, we compare the experimental results with \textit{ab-initio} calculations to investigate the microscopic origin of pressure evolution of magnetism in \cri . The structural optimization we performed results in a NN Cr-Cr distance of 4.04~\AA, which is in a reasonable agreement with the experimental value of 3.96~\AA.~\cite{mcguire2015} The Cr-I-Cr bond angle was found to be 95$^\circ$ in the calculations and 93.3$^\circ$ in experiment. Overall, our DFT calculations result in a good description of the intra-layer structural parameters. In case of the inter-layer spacing, the agreement is less good: 7.3~\AA~against 6.602~\AA~ in experiment. This disagreement originates from neglecting the van-der-Waals interactions which govern the coupling between the layers. The estimates were shown to improve by employing vdW-corrected functionals.~\cite{mcguire2015} Due to the different quality of the predicted estimates for in-plane and out-of-plane constants, one can expect the results for in-plane strain to be more reliable. However, these differences should not affect the overall trends in the changes of the exchange interactions as a function of strain.

\begin{figure}[!h]
\includegraphics[width=\columnwidth]{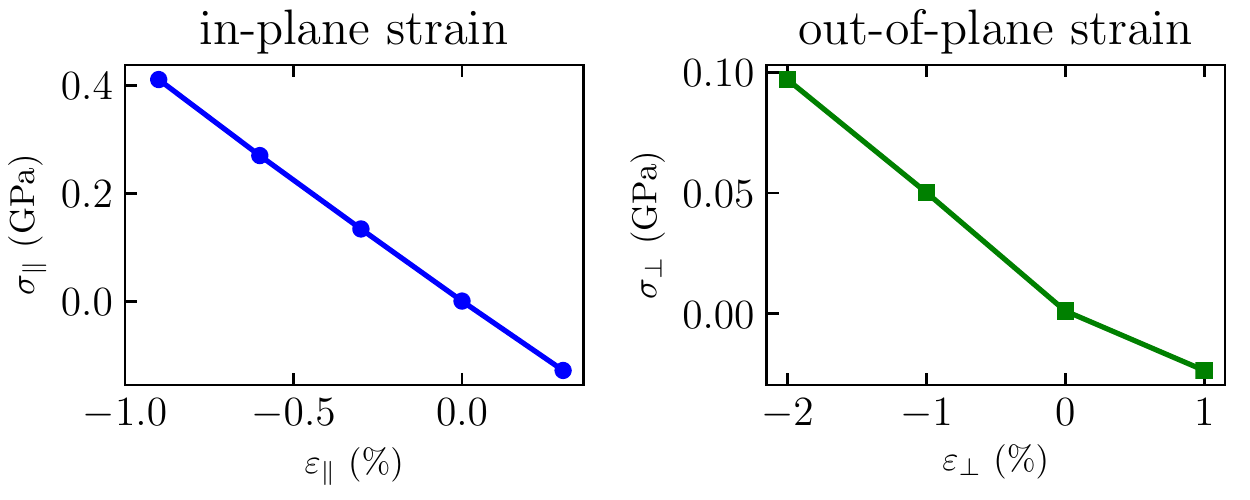} 
\caption{Calculated stress ($\sigma$) for different values of strain ($\epsilon$) applied along the same direction.\label{eps-sigma-GPa}}
\end{figure}

Fig.~\ref{eps-sigma-GPa} shows the calculated stress for a given value of applied strain applied within the plane of the atomic layers and perpendicular to it. Here, strain is $\epsilon_i=dL_i/L_i$. It can be clearly seen that compressing the lattice along the $c$ direction is associated with considerably smaller stress as compared to compression within the $ab$ plane, which is an expected result for systems consisting of weakly coupled layers. 

\begin{figure}[!h]
\includegraphics[width=\columnwidth]{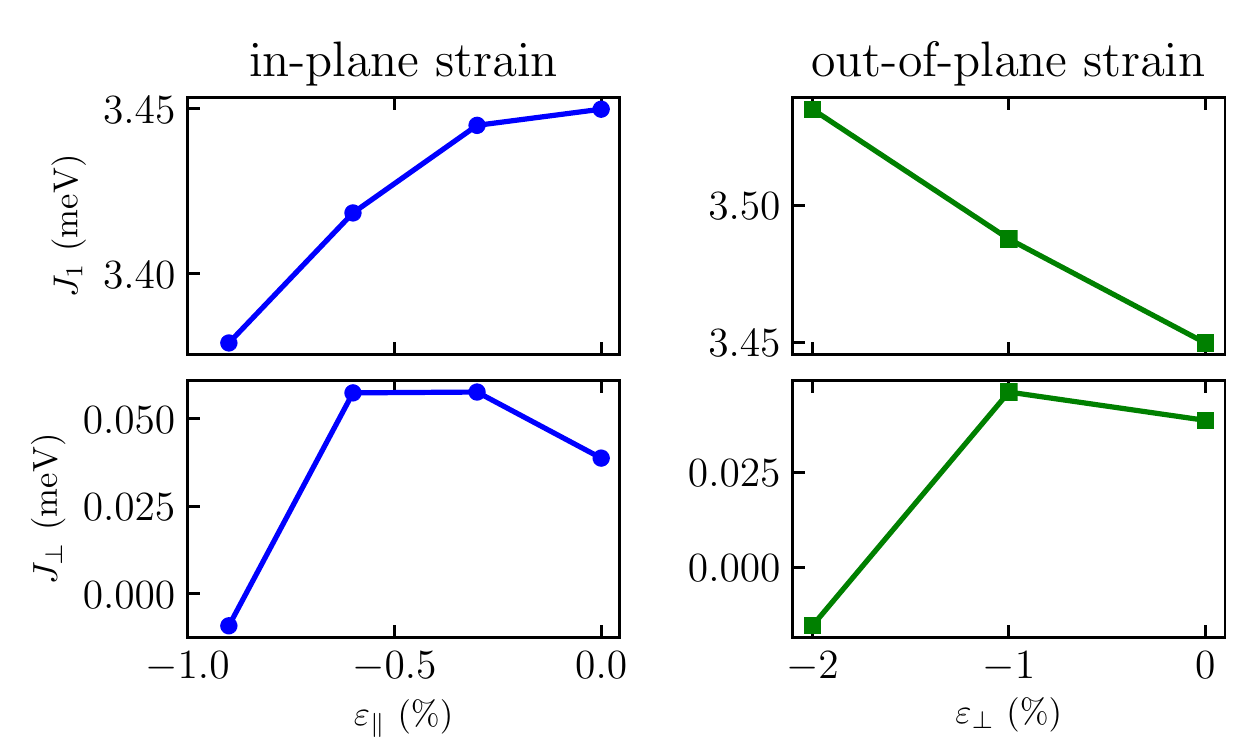} 
\caption{Calculated nearest-neighbour ($J_1$) and first out-of-plane exchange interaction ($J_{\perp}$) as a function of strain. Positive sign of the magnetic interaction corresponds to the FM coupling.}
\label{eps-Jij}
\end{figure}

Next, we constructed supercells of the obtained crystal structures and extracted the most important exchange parameters.
The results shown in Fig.~\ref{eps-Jij} reveal that compressive in-plane strain tends to decrease the values of the dominant $J_1$ interaction while applying the stress along the $c$ direction leads to a weak increase. 
Inter-layer coupling $J_{\perp}$ responds to both in-plane and out-of-plane compression in a similar way. Being relatively small, $J_{\perp}$ is shown to be quite sensitive to strain and is even able to change its sign from FM to AFM for a sufficiently strong compression. Similarly, numerical calculations on the strain dependence of the exchange parameters in bilayer \cri{} also imply a transition of the magnetic inter-layer coupling from AFM to FM at a compressive in-plane strain of $\sim 1\,\%$.~\cite{leon2020}

Strain dependence of the exchange parameters and the $T_c$ in monolayers of CrX$_3$ (X=$\{$Cl,~Br,~I$\}$) was studied in Ref.~\onlinecite{webster2018}.
It was shown that compressive (in-plane) strain leads to the decrease of the nearest-neighbour (NN) Cr-Cr exchange interaction in the whole series of Cr$X_3$ monolayers.
For sufficiently large compression, the sign of the coupling even switches from ferromagnetic to antiferromagnetic.
In Ref.~\onlinecite{dupont2020} the physical origin of this change was explained in detail.
In these systems, the Cr atoms are surrounded by edge-sharing octahedra and the exchange interactions are mediated by Cr-X-Cr networks.
In this situation, there are two competing contributions to the NN exchange coupling.~\cite{anderson1959, kanamori1959, goodenough1958, besbes2019, kashin2020}
First, there is a FM superexchange interaction between half-filled $t_{2g}$-like orbitals on one site and nominally empty $e_g$-like states on the neighbouring Cr atom.
This coupling is opposed by the AFM exchange between half-filled $t_{2g}$ orbitals which is governed by two mechanisms -- the superexchange via a single halide $p$ orbital and \textit{direct} kinetic exchange between the $t_{2g}$ orbitals pointing towards each other.
As one compresses the lattice, the balance between these contributions shifts and the AFM kinetic term starts to dominate, changing the sign of the total exchange integral.

Our results for the strain dependence of the NN $J_1$ coupling fall within the same scenario. As for $J_{\perp}$, this coupling is rather weak and is likely to be mediated by higher-order superexchange processes involving halide orbitals.~\cite{sivadas2018}
It is likely that a decrease of inter-layer spacing tends to increase the overlap between wavefunctions involved in the dominating exchange process. In fact, this interaction is known to be strongly dependent on the layer stacking.~\cite{song2019, li2019} In our calculations we assumed the stacking not to change, which is a reasonable approximation for such small values of strain as we are dealing with here.

\section{Summary}

In summary, our high-resolution thermal expansion studies are used to elucidate and quantify magnetoelastic coupling and uniaxial pressure effects in CrI$_3$. Thermal expansion anomalies \ac\ and \aab\ at the ferromagnetic ordering temperature evidence magnetoelastic coupling and imply positive (negative) pressure dependencies $\partial T_{\rm C}/\partial p_{\rm i}$ for pressure applied along $c$ ($ab$). While uniaxial pressure applied in-plane yields $\partial T_{\rm C}/\partial p_{\rm ab}\simeq -0.4$~K/GPa, the pressure dependence on $p\|c$ is positive and about four times larger. The response of ferromagnetism in \cri{} to applied stress is also confirmed by numerical calculations showing that the dominating NN coupling increases as the lattice is compressed along the $c$ direction, whereas negative in-plane strain leads to a reduction of $J_1$. In contrast, inter-layer exchange $J_{\perp}$ is shown to initially increase and afterwards decrease upon the application of both out-of-plane and in-plane compression. Therefore, the experimentally observed uniaxial pressure dependencies of \tc\ follow that of $J_1$ (but not of $J_{\perp}$) which confirms the dominant role of $J_1$ for the evolution of long-range ferromagnetic order in CrI$_3$. Overall, the results show that macroscopic length changes can be used to investigate microscopic parameters and in particular provide experimental values for \textit{uniaxial} pressure dependencies in quasi-two-dimensional systems.

\begin{acknowledgements}
We acknowledge financial support by BMBF via the project SpinFun (13XP5088) and by Deutsche Forschungsgemeinschaft (DFG) under Germany’s Excellence Strategy EXC2181/1-390900948 (the Heidelberg STRUCTURES Excellence Cluster) and through project KL 1824/13-1. Support by the Ministry of Education and Science of the Russian Federation within the framework of the governmental program of Megagrants 2020-220-08-6358 is acknowledged. We also acknowledge support by the European Union’s Horizon 2020 Research and Innovation Programme, under Grant Agreement no. 824109 (European Microkelvin Platform). M.A.H. and Y.O.K. acknowledge the financial support from the Swedish Research Council (VR) under the project no. 2018-05393 and 2019-03569, respectively. The DFT calculations were performed using the resources provided by the Swedish National Infrastructure for Computing (SNIC) at the National Supercomputing Centre (NSC).
\end{acknowledgements}

\bibliography{Bib_CrI3_v1}

\begin{thebibliography}{55}
\expandafter\ifx\csname natexlab\endcsname\relax\def\natexlab#1{#1}\fi
\expandafter\ifx\csname bibnamefont\endcsname\relax
  \def\bibnamefont#1{#1}\fi
\expandafter\ifx\csname bibfnamefont\endcsname\relax
  \def\bibfnamefont#1{#1}\fi
\expandafter\ifx\csname citenamefont\endcsname\relax
  \def\citenamefont#1{#1}\fi
\expandafter\ifx\csname url\endcsname\relax
  \def\url#1{\texttt{#1}}\fi
\expandafter\ifx\csname urlprefix\endcsname\relax\def\urlprefix{URL }\fi
\providecommand{\bibinfo}[2]{#2}
\providecommand{\eprint}[2][]{\url{#2}}

\bibitem[{\citenamefont{Huang et~al.}(2017)\citenamefont{Huang, Clark,
  Navarro-Moratalla, Klein, Cheng, Seyler, Zhong, Schmidgall, McGuire, Cobden
  et~al.}}]{huang2017}
\bibinfo{author}{\bibfnamefont{B.}~\bibnamefont{Huang}},
  \bibinfo{author}{\bibfnamefont{G.}~\bibnamefont{Clark}},
  \bibinfo{author}{\bibfnamefont{E.}~\bibnamefont{Navarro-Moratalla}},
  \bibinfo{author}{\bibfnamefont{D.~R.} \bibnamefont{Klein}},
  \bibinfo{author}{\bibfnamefont{R.}~\bibnamefont{Cheng}},
  \bibinfo{author}{\bibfnamefont{K.~L.} \bibnamefont{Seyler}},
  \bibinfo{author}{\bibfnamefont{D.}~\bibnamefont{Zhong}},
  \bibinfo{author}{\bibfnamefont{E.}~\bibnamefont{Schmidgall}},
  \bibinfo{author}{\bibfnamefont{M.~A.} \bibnamefont{McGuire}},
  \bibinfo{author}{\bibfnamefont{D.~H.} \bibnamefont{Cobden}},
  \bibnamefont{et~al.}, \bibinfo{journal}{Nature}
  \textbf{\bibinfo{volume}{546}}, \bibinfo{pages}{270} (\bibinfo{year}{2017}).

\bibitem[{\citenamefont{Gong et~al.}(2017)\citenamefont{Gong, Li, Li, Ji,
  Stern, Xia, Cao, Bao, Wang, Wang et~al.}}]{gong2017}
\bibinfo{author}{\bibfnamefont{C.}~\bibnamefont{Gong}},
  \bibinfo{author}{\bibfnamefont{L.}~\bibnamefont{Li}},
  \bibinfo{author}{\bibfnamefont{Z.}~\bibnamefont{Li}},
  \bibinfo{author}{\bibfnamefont{H.}~\bibnamefont{Ji}},
  \bibinfo{author}{\bibfnamefont{A.}~\bibnamefont{Stern}},
  \bibinfo{author}{\bibfnamefont{Y.}~\bibnamefont{Xia}},
  \bibinfo{author}{\bibfnamefont{T.}~\bibnamefont{Cao}},
  \bibinfo{author}{\bibfnamefont{W.}~\bibnamefont{Bao}},
  \bibinfo{author}{\bibfnamefont{C.}~\bibnamefont{Wang}},
  \bibinfo{author}{\bibfnamefont{Y.}~\bibnamefont{Wang}}, \bibnamefont{et~al.},
  \bibinfo{journal}{Nature} \textbf{\bibinfo{volume}{546}},
  \bibinfo{pages}{265} (\bibinfo{year}{2017}).

\bibitem[{\citenamefont{O'Hara et~al.}(2018)\citenamefont{O'Hara, Zhu, Trout,
  Ahmed, Luo, Lee, Brenner, Rajan, Gupta, McComb et~al.}}]{OHara2018}
\bibinfo{author}{\bibfnamefont{D.~J.} \bibnamefont{O'Hara}},
  \bibinfo{author}{\bibfnamefont{T.}~\bibnamefont{Zhu}},
  \bibinfo{author}{\bibfnamefont{A.~H.} \bibnamefont{Trout}},
  \bibinfo{author}{\bibfnamefont{A.~S.} \bibnamefont{Ahmed}},
  \bibinfo{author}{\bibfnamefont{Y.~K.} \bibnamefont{Luo}},
  \bibinfo{author}{\bibfnamefont{C.~H.} \bibnamefont{Lee}},
  \bibinfo{author}{\bibfnamefont{M.~R.} \bibnamefont{Brenner}},
  \bibinfo{author}{\bibfnamefont{S.}~\bibnamefont{Rajan}},
  \bibinfo{author}{\bibfnamefont{J.~A.} \bibnamefont{Gupta}},
  \bibinfo{author}{\bibfnamefont{D.~W.} \bibnamefont{McComb}},
  \bibnamefont{et~al.}, \bibinfo{journal}{Nano Lett.}
  \textbf{\bibinfo{volume}{18}}, \bibinfo{pages}{3125} (\bibinfo{year}{2018}).

\bibitem[{\citenamefont{Sun et~al.}(2020)\citenamefont{Sun, Li, Wang, Sui,
  Zhang, Wang, Liu, Li, Feng, Zhong et~al.}}]{sun2020}
\bibinfo{author}{\bibfnamefont{X.}~\bibnamefont{Sun}},
  \bibinfo{author}{\bibfnamefont{W.}~\bibnamefont{Li}},
  \bibinfo{author}{\bibfnamefont{X.}~\bibnamefont{Wang}},
  \bibinfo{author}{\bibfnamefont{Q.}~\bibnamefont{Sui}},
  \bibinfo{author}{\bibfnamefont{T.}~\bibnamefont{Zhang}},
  \bibinfo{author}{\bibfnamefont{Z.}~\bibnamefont{Wang}},
  \bibinfo{author}{\bibfnamefont{L.}~\bibnamefont{Liu}},
  \bibinfo{author}{\bibfnamefont{D.}~\bibnamefont{Li}},
  \bibinfo{author}{\bibfnamefont{S.}~\bibnamefont{Feng}},
  \bibinfo{author}{\bibfnamefont{S.}~\bibnamefont{Zhong}},
  \bibnamefont{et~al.}, \bibinfo{journal}{Nano Res.}
  \textbf{\bibinfo{volume}{13}}, \bibinfo{pages}{3358} (\bibinfo{year}{2020}).

\bibitem[{\citenamefont{Jenjeti et~al.}(2018)\citenamefont{Jenjeti, Kumar,
  Austeria, and Sampath}}]{jenjeti2018}
\bibinfo{author}{\bibfnamefont{R.~B.} \bibnamefont{Jenjeti}},
  \bibinfo{author}{\bibfnamefont{R.}~\bibnamefont{Kumar}},
  \bibinfo{author}{\bibfnamefont{M.~P.} \bibnamefont{Austeria}},
  \bibnamefont{and} \bibinfo{author}{\bibfnamefont{S.}~\bibnamefont{Sampath}},
  \bibinfo{journal}{Sci. Rep.} \textbf{\bibinfo{volume}{8}},
  \bibinfo{pages}{8586} (\bibinfo{year}{2018}).

\bibitem[{\citenamefont{Idzuchi et~al.}(2019)\citenamefont{Idzuchi,
  Llacsahuanga{ }Allcca, Pan, Tanigaki, and Chen}}]{idzuchi2019}
\bibinfo{author}{\bibfnamefont{H.}~\bibnamefont{Idzuchi}},
  \bibinfo{author}{\bibfnamefont{A.~E.} \bibnamefont{Llacsahuanga{ }Allcca}},
  \bibinfo{author}{\bibfnamefont{X.~C.} \bibnamefont{Pan}},
  \bibinfo{author}{\bibfnamefont{K.}~\bibnamefont{Tanigaki}}, \bibnamefont{and}
  \bibinfo{author}{\bibfnamefont{Y.~P.} \bibnamefont{Chen}},
  \bibinfo{journal}{Appl. Phys. Lett.} \textbf{\bibinfo{volume}{115}},
  \bibinfo{pages}{232403} (\bibinfo{year}{2019}).

\bibitem[{\citenamefont{Mervin and Wagner}(1966)}]{mervinwagner}
\bibinfo{author}{\bibfnamefont{N.~D.} \bibnamefont{Mervin}} \bibnamefont{and}
  \bibinfo{author}{\bibfnamefont{H.}~\bibnamefont{Wagner}},
  \bibinfo{journal}{Phys. Rev. Lett.} \textbf{\bibinfo{volume}{17}},
  \bibinfo{pages}{1133} (\bibinfo{year}{1966}).

\bibitem[{\citenamefont{Besbes et~al.}(2019)\citenamefont{Besbes, Nikolaev,
  Meskini, and Solovyev}}]{besbes2019}
\bibinfo{author}{\bibfnamefont{O.}~\bibnamefont{Besbes}},
  \bibinfo{author}{\bibfnamefont{S.}~\bibnamefont{Nikolaev}},
  \bibinfo{author}{\bibfnamefont{N.}~\bibnamefont{Meskini}}, \bibnamefont{and}
  \bibinfo{author}{\bibfnamefont{I.}~\bibnamefont{Solovyev}},
  \bibinfo{journal}{Phys. Rev. B} \textbf{\bibinfo{volume}{99}},
  \bibinfo{pages}{104432} (\bibinfo{year}{2019}).

\bibitem[{\citenamefont{Soriano et~al.}(2020)\citenamefont{Soriano, Katsnelson,
  and Fernández-Rossier}}]{soriano2020}
\bibinfo{author}{\bibfnamefont{D.}~\bibnamefont{Soriano}},
  \bibinfo{author}{\bibfnamefont{M.~I.} \bibnamefont{Katsnelson}},
  \bibnamefont{and}
  \bibinfo{author}{\bibfnamefont{J.}~\bibnamefont{Fernández-Rossier}},
  \bibinfo{journal}{Nano Lett.} \textbf{\bibinfo{volume}{20}},
  \bibinfo{pages}{6225} (\bibinfo{year}{2020}).

\bibitem[{\citenamefont{Chen et~al.}(2020)\citenamefont{Chen, Chung, Chen,
  Duan, Schneidewind, Radelytskyi, Voneshen, Ewings, Stone, Kolesnikov
  et~al.}}]{chen2020}
\bibinfo{author}{\bibfnamefont{L.}~\bibnamefont{Chen}},
  \bibinfo{author}{\bibfnamefont{J.-H.} \bibnamefont{Chung}},
  \bibinfo{author}{\bibfnamefont{T.}~\bibnamefont{Chen}},
  \bibinfo{author}{\bibfnamefont{C.}~\bibnamefont{Duan}},
  \bibinfo{author}{\bibfnamefont{A.}~\bibnamefont{Schneidewind}},
  \bibinfo{author}{\bibfnamefont{I.}~\bibnamefont{Radelytskyi}},
  \bibinfo{author}{\bibfnamefont{D.~J.} \bibnamefont{Voneshen}},
  \bibinfo{author}{\bibfnamefont{R.~A.} \bibnamefont{Ewings}},
  \bibinfo{author}{\bibfnamefont{M.~B.} \bibnamefont{Stone}},
  \bibinfo{author}{\bibfnamefont{A.~I.} \bibnamefont{Kolesnikov}},
  \bibnamefont{et~al.}, \bibinfo{journal}{Phys. Rev. B}
  \textbf{\bibinfo{volume}{101}}, \bibinfo{pages}{134418}
  (\bibinfo{year}{2020}).

\bibitem[{\citenamefont{Chen et~al.}(2018)\citenamefont{Chen, Chung, Gao, Chen,
  Stone, Kolesnikov, Huang, and Dai}}]{chen2018}
\bibinfo{author}{\bibfnamefont{L.}~\bibnamefont{Chen}},
  \bibinfo{author}{\bibfnamefont{J.-H.} \bibnamefont{Chung}},
  \bibinfo{author}{\bibfnamefont{B.}~\bibnamefont{Gao}},
  \bibinfo{author}{\bibfnamefont{T.}~\bibnamefont{Chen}},
  \bibinfo{author}{\bibfnamefont{M.~B.} \bibnamefont{Stone}},
  \bibinfo{author}{\bibfnamefont{A.~I.} \bibnamefont{Kolesnikov}},
  \bibinfo{author}{\bibfnamefont{Q.}~\bibnamefont{Huang}}, \bibnamefont{and}
  \bibinfo{author}{\bibfnamefont{P.}~\bibnamefont{Dai}},
  \bibinfo{journal}{Phys. Rev. X} \textbf{\bibinfo{volume}{8}},
  \bibinfo{pages}{041028} (\bibinfo{year}{2018}).

\bibitem[{\citenamefont{Xu et~al.}(2018)\citenamefont{Xu, Feng, Xiang, and
  Bellaiche}}]{xu2018}
\bibinfo{author}{\bibfnamefont{C.}~\bibnamefont{Xu}},
  \bibinfo{author}{\bibfnamefont{J.}~\bibnamefont{Feng}},
  \bibinfo{author}{\bibfnamefont{H.}~\bibnamefont{Xiang}}, \bibnamefont{and}
  \bibinfo{author}{\bibfnamefont{L.}~\bibnamefont{Bellaiche}},
  \bibinfo{journal}{npj Comput. Mater.} \textbf{\bibinfo{volume}{4}},
  \bibinfo{pages}{57} (\bibinfo{year}{2018}).

\bibitem[{\citenamefont{Lee et~al.}(2020)\citenamefont{Lee, Utermohlen, Weber,
  Hwang, Zhang, van Tol, Goldberger, Trivedi, and Hammel}}]{lee2020}
\bibinfo{author}{\bibfnamefont{I.}~\bibnamefont{Lee}},
  \bibinfo{author}{\bibfnamefont{F.~G.} \bibnamefont{Utermohlen}},
  \bibinfo{author}{\bibfnamefont{D.}~\bibnamefont{Weber}},
  \bibinfo{author}{\bibfnamefont{K.}~\bibnamefont{Hwang}},
  \bibinfo{author}{\bibfnamefont{C.}~\bibnamefont{Zhang}},
  \bibinfo{author}{\bibfnamefont{J.}~\bibnamefont{van Tol}},
  \bibinfo{author}{\bibfnamefont{J.}~\bibnamefont{Goldberger}},
  \bibinfo{author}{\bibfnamefont{N.}~\bibnamefont{Trivedi}}, \bibnamefont{and}
  \bibinfo{author}{\bibfnamefont{P.~C.} \bibnamefont{Hammel}},
  \bibinfo{journal}{Phys. Rev. Lett.} \textbf{\bibinfo{volume}{124}},
  \bibinfo{pages}{017201} (\bibinfo{year}{2020}).

\bibitem[{\citenamefont{Kartsev et~al.}(2020)\citenamefont{Kartsev, Augustin,
  Evans, Novoselov, and Santos}}]{kartsev2020}
\bibinfo{author}{\bibfnamefont{A.}~\bibnamefont{Kartsev}},
  \bibinfo{author}{\bibfnamefont{M.}~\bibnamefont{Augustin}},
  \bibinfo{author}{\bibfnamefont{R.~F.~L.} \bibnamefont{Evans}},
  \bibinfo{author}{\bibfnamefont{K.~S.} \bibnamefont{Novoselov}},
  \bibnamefont{and} \bibinfo{author}{\bibfnamefont{E.~J.~G.}
  \bibnamefont{Santos}}, \bibinfo{journal}{npj Comput. Mater.}
  \textbf{\bibinfo{volume}{6}}, \bibinfo{pages}{150} (\bibinfo{year}{2020}).

\bibitem[{\citenamefont{McGuire et~al.}(2015)\citenamefont{McGuire, Dixit,
  Cooper, and Sales}}]{mcguire2015}
\bibinfo{author}{\bibfnamefont{M.~A.} \bibnamefont{McGuire}},
  \bibinfo{author}{\bibfnamefont{H.}~\bibnamefont{Dixit}},
  \bibinfo{author}{\bibfnamefont{V.~R.} \bibnamefont{Cooper}},
  \bibnamefont{and} \bibinfo{author}{\bibfnamefont{B.~C.} \bibnamefont{Sales}},
  \bibinfo{journal}{Chem. Mater.} \textbf{\bibinfo{volume}{27}},
  \bibinfo{pages}{612} (\bibinfo{year}{2015}).

\bibitem[{\citenamefont{Ubrig et~al.}(2020)\citenamefont{Ubrig, Wang, Teyssier,
  Taniguchi, Watanabe, Giannini, Morpurgo, and Gilbertini}}]{ubrig2019}
\bibinfo{author}{\bibfnamefont{N.}~\bibnamefont{Ubrig}},
  \bibinfo{author}{\bibfnamefont{Z.}~\bibnamefont{Wang}},
  \bibinfo{author}{\bibfnamefont{J.}~\bibnamefont{Teyssier}},
  \bibinfo{author}{\bibfnamefont{T.}~\bibnamefont{Taniguchi}},
  \bibinfo{author}{\bibfnamefont{K.}~\bibnamefont{Watanabe}},
  \bibinfo{author}{\bibfnamefont{E.}~\bibnamefont{Giannini}},
  \bibinfo{author}{\bibfnamefont{A.~F.} \bibnamefont{Morpurgo}},
  \bibnamefont{and}
  \bibinfo{author}{\bibfnamefont{M.}~\bibnamefont{Gilbertini}},
  \bibinfo{journal}{2D Mater.} \textbf{\bibinfo{volume}{7}},
  \bibinfo{pages}{015007} (\bibinfo{year}{2020}).

\bibitem[{\citenamefont{Sivadas et~al.}(2018)\citenamefont{Sivadas, Okamoto,
  Xu, Fennie, and Xiao}}]{sivadas2018}
\bibinfo{author}{\bibfnamefont{N.}~\bibnamefont{Sivadas}},
  \bibinfo{author}{\bibfnamefont{S.}~\bibnamefont{Okamoto}},
  \bibinfo{author}{\bibfnamefont{X.}~\bibnamefont{Xu}},
  \bibinfo{author}{\bibfnamefont{C.~J.} \bibnamefont{Fennie}},
  \bibnamefont{and} \bibinfo{author}{\bibfnamefont{D.}~\bibnamefont{Xiao}},
  \bibinfo{journal}{Nano Lett.} \textbf{\bibinfo{volume}{18}},
  \bibinfo{pages}{7658} (\bibinfo{year}{2018}).

\bibitem[{\citenamefont{Küchler et~al.}(2012)\citenamefont{Küchler, Bauer,
  Brando, and Steglich}}]{kuechler2012}
\bibinfo{author}{\bibfnamefont{R.}~\bibnamefont{Küchler}},
  \bibinfo{author}{\bibfnamefont{T.}~\bibnamefont{Bauer}},
  \bibinfo{author}{\bibfnamefont{M.}~\bibnamefont{Brando}}, \bibnamefont{and}
  \bibinfo{author}{\bibfnamefont{F.}~\bibnamefont{Steglich}},
  \bibinfo{journal}{Review of Scientific Instruments}
  \textbf{\bibinfo{volume}{83}}, \bibinfo{pages}{095102}
  (\bibinfo{year}{2012}).

\bibitem[{\citenamefont{Werner et~al.}(2017)\citenamefont{Werner, Hergett,
  Gertig, Park, Koo, and Klingeler}}]{werner2017}
\bibinfo{author}{\bibfnamefont{J.}~\bibnamefont{Werner}},
  \bibinfo{author}{\bibfnamefont{W.}~\bibnamefont{Hergett}},
  \bibinfo{author}{\bibfnamefont{M.}~\bibnamefont{Gertig}},
  \bibinfo{author}{\bibfnamefont{J.}~\bibnamefont{Park}},
  \bibinfo{author}{\bibfnamefont{C.}~\bibnamefont{Koo}}, \bibnamefont{and}
  \bibinfo{author}{\bibfnamefont{R.}~\bibnamefont{Klingeler}},
  \bibinfo{journal}{Phys. Rev. B} \textbf{\bibinfo{volume}{95}},
  \bibinfo{pages}{214414} (\bibinfo{year}{2017}).

\bibitem[{\citenamefont{Perdew et~al.}(1996)\citenamefont{Perdew, Burke, and
  Ernzerhof}}]{gga-pbe}
\bibinfo{author}{\bibfnamefont{J.~P.} \bibnamefont{Perdew}},
  \bibinfo{author}{\bibfnamefont{K.}~\bibnamefont{Burke}}, \bibnamefont{and}
  \bibinfo{author}{\bibfnamefont{M.}~\bibnamefont{Ernzerhof}},
  \bibinfo{journal}{Phys. Rev. Lett.} \textbf{\bibinfo{volume}{77}},
  \bibinfo{pages}{3865} (\bibinfo{year}{1996}).

\bibitem[{\citenamefont{Kresse and Joubert}(1999)}]{paw}
\bibinfo{author}{\bibfnamefont{G.}~\bibnamefont{Kresse}} \bibnamefont{and}
  \bibinfo{author}{\bibfnamefont{D.}~\bibnamefont{Joubert}},
  \bibinfo{journal}{Phys. Rev. B} \textbf{\bibinfo{volume}{59}},
  \bibinfo{pages}{1758} (\bibinfo{year}{1999}).

\bibitem[{\citenamefont{Kresse and Furthmüller}(1996)}]{vasp}
\bibinfo{author}{\bibfnamefont{G.}~\bibnamefont{Kresse}} \bibnamefont{and}
  \bibinfo{author}{\bibfnamefont{J.}~\bibnamefont{Furthmüller}},
  \bibinfo{journal}{Comput. Mater. Sci.} \textbf{\bibinfo{volume}{6}},
  \bibinfo{pages}{15 } (\bibinfo{year}{1996}).

\bibitem[{\citenamefont{Lu et~al.}(2019)\citenamefont{Lu, Fei, and
  Yang}}]{lu2019}
\bibinfo{author}{\bibfnamefont{X.}~\bibnamefont{Lu}},
  \bibinfo{author}{\bibfnamefont{R.}~\bibnamefont{Fei}}, \bibnamefont{and}
  \bibinfo{author}{\bibfnamefont{L.}~\bibnamefont{Yang}},
  \bibinfo{journal}{Phys. Rev. B} \textbf{\bibinfo{volume}{100}},
  \bibinfo{pages}{205409} (\bibinfo{year}{2019}).

\bibitem[{\citenamefont{Lado and Fern\^{a}ndez-Rossier}(2017)}]{lado2017}
\bibinfo{author}{\bibfnamefont{J.~L.} \bibnamefont{Lado}} \bibnamefont{and}
  \bibinfo{author}{\bibfnamefont{F.}~\bibnamefont{Fern\^{a}ndez-Rossier}},
  \bibinfo{journal}{2D Mater.} \textbf{\bibinfo{volume}{4}},
  \bibinfo{pages}{035002} (\bibinfo{year}{2017}).

\bibitem[{\citenamefont{Webster and Yan}(2018)}]{webster2018}
\bibinfo{author}{\bibfnamefont{L.}~\bibnamefont{Webster}} \bibnamefont{and}
  \bibinfo{author}{\bibfnamefont{J.}~\bibnamefont{Yan}},
  \bibinfo{journal}{Phys. Rev. B} \textbf{\bibinfo{volume}{98}},
  \bibinfo{pages}{144411} (\bibinfo{year}{2018}).

\bibitem[{\citenamefont{Xu et~al.}(2020)\citenamefont{Xu, Feng, Prokhorenko,
  Nahas, Xiang, and Bellaiche}}]{xu2020}
\bibinfo{author}{\bibfnamefont{C.}~\bibnamefont{Xu}},
  \bibinfo{author}{\bibfnamefont{J.}~\bibnamefont{Feng}},
  \bibinfo{author}{\bibfnamefont{S.}~\bibnamefont{Prokhorenko}},
  \bibinfo{author}{\bibfnamefont{Y.}~\bibnamefont{Nahas}},
  \bibinfo{author}{\bibfnamefont{H.}~\bibnamefont{Xiang}}, \bibnamefont{and}
  \bibinfo{author}{\bibfnamefont{L.}~\bibnamefont{Bellaiche}},
  \bibinfo{journal}{Phys. Rev. B} \textbf{\bibinfo{volume}{101}},
  \bibinfo{pages}{060404} (\bibinfo{year}{2020}).

\bibitem[{\citenamefont{Olsen}(2019)}]{olsen2019}
\bibinfo{author}{\bibfnamefont{T.}~\bibnamefont{Olsen}}, \bibinfo{journal}{M R
  S Commun.} \textbf{\bibinfo{volume}{9}}, \bibinfo{pages}{1142–1150}
  (\bibinfo{year}{2019}).

\bibitem[{\citenamefont{Xue et~al.}(2019)\citenamefont{Xue, Hou, Wang, and
  Wu}}]{xue2019}
\bibinfo{author}{\bibfnamefont{F.}~\bibnamefont{Xue}},
  \bibinfo{author}{\bibfnamefont{Y.}~\bibnamefont{Hou}},
  \bibinfo{author}{\bibfnamefont{Z.}~\bibnamefont{Wang}}, \bibnamefont{and}
  \bibinfo{author}{\bibfnamefont{R.}~\bibnamefont{Wu}}, \bibinfo{journal}{Phys.
  Rev. B} \textbf{\bibinfo{volume}{100}}, \bibinfo{pages}{224429}
  (\bibinfo{year}{2019}).

\bibitem[{\citenamefont{Zhang et~al.}(2015)\citenamefont{Zhang, Qu, Zhu, and
  Lam}}]{zhang2015}
\bibinfo{author}{\bibfnamefont{W.}~\bibnamefont{Zhang}},
  \bibinfo{author}{\bibfnamefont{Q.}~\bibnamefont{Qu}},
  \bibinfo{author}{\bibfnamefont{P.}~\bibnamefont{Zhu}}, \bibnamefont{and}
  \bibinfo{author}{\bibfnamefont{C.}~\bibnamefont{Lam}}, \bibinfo{journal}{J.
  Mater Chem. C} \textbf{\bibinfo{volume}{3}}, \bibinfo{pages}{12457}
  (\bibinfo{year}{2015}).

\bibitem[{\citenamefont{Pizzochero and Yazyev}(2020)}]{pizzochero2020}
\bibinfo{author}{\bibfnamefont{M.}~\bibnamefont{Pizzochero}} \bibnamefont{and}
  \bibinfo{author}{\bibfnamefont{O.~V.} \bibnamefont{Yazyev}},
  \bibinfo{journal}{J. Phys. Chem. C} \textbf{\bibinfo{volume}{124}},
  \bibinfo{pages}{7585} (\bibinfo{year}{2020}).

\bibitem[{\citenamefont{Niu et~al.}(2020)\citenamefont{Niu, Su, Francisco,
  Ghosh, Kargar, Huang, Lohmann, Li, Xu, Taniguchi et~al.}}]{niu2020}
\bibinfo{author}{\bibfnamefont{B.}~\bibnamefont{Niu}},
  \bibinfo{author}{\bibfnamefont{T.}~\bibnamefont{Su}},
  \bibinfo{author}{\bibfnamefont{B.~A.} \bibnamefont{Francisco}},
  \bibinfo{author}{\bibfnamefont{S.}~\bibnamefont{Ghosh}},
  \bibinfo{author}{\bibfnamefont{F.}~\bibnamefont{Kargar}},
  \bibinfo{author}{\bibfnamefont{X.}~\bibnamefont{Huang}},
  \bibinfo{author}{\bibfnamefont{M.}~\bibnamefont{Lohmann}},
  \bibinfo{author}{\bibfnamefont{J.}~\bibnamefont{Li}},
  \bibinfo{author}{\bibfnamefont{Y.}~\bibnamefont{Xu}},
  \bibinfo{author}{\bibfnamefont{T.}~\bibnamefont{Taniguchi}},
  \bibnamefont{et~al.}, \bibinfo{journal}{Nano Lett.}
  \textbf{\bibinfo{volume}{20}}, \bibinfo{pages}{553} (\bibinfo{year}{2020}).

\bibitem[{\citenamefont{Li et~al.}(2020)\citenamefont{Li, Ye, Luo, Ye, Kim,
  Yang, Tian, Li, Lei, Tsen et~al.}}]{li2020}
\bibinfo{author}{\bibfnamefont{S.}~\bibnamefont{Li}},
  \bibinfo{author}{\bibfnamefont{Z.}~\bibnamefont{Ye}},
  \bibinfo{author}{\bibfnamefont{X.}~\bibnamefont{Luo}},
  \bibinfo{author}{\bibfnamefont{G.}~\bibnamefont{Ye}},
  \bibinfo{author}{\bibfnamefont{H.~H.} \bibnamefont{Kim}},
  \bibinfo{author}{\bibfnamefont{B.}~\bibnamefont{Yang}},
  \bibinfo{author}{\bibfnamefont{S.}~\bibnamefont{Tian}},
  \bibinfo{author}{\bibfnamefont{C.}~\bibnamefont{Li}},
  \bibinfo{author}{\bibfnamefont{H.}~\bibnamefont{Lei}},
  \bibinfo{author}{\bibfnamefont{A.}~\bibnamefont{Tsen}}, \bibnamefont{et~al.},
  \bibinfo{journal}{Phys. Rev. X} \textbf{\bibinfo{volume}{10}},
  \bibinfo{pages}{011075} (\bibinfo{year}{2020}).

\bibitem[{\citenamefont{McCreary et~al.}(2020)\citenamefont{McCreary, Mai,
  Utermohlen, Simpson, Garrity, Feng, Shcherbakov, Zhu, Hu, Weber
  et~al.}}]{mccreary2020}
\bibinfo{author}{\bibfnamefont{A.}~\bibnamefont{McCreary}},
  \bibinfo{author}{\bibfnamefont{T.~T.} \bibnamefont{Mai}},
  \bibinfo{author}{\bibfnamefont{F.~G.} \bibnamefont{Utermohlen}},
  \bibinfo{author}{\bibfnamefont{J.~R.} \bibnamefont{Simpson}},
  \bibinfo{author}{\bibfnamefont{K.~F.} \bibnamefont{Garrity}},
  \bibinfo{author}{\bibfnamefont{X.}~\bibnamefont{Feng}},
  \bibinfo{author}{\bibfnamefont{D.}~\bibnamefont{Shcherbakov}},
  \bibinfo{author}{\bibfnamefont{Y.}~\bibnamefont{Zhu}},
  \bibinfo{author}{\bibfnamefont{J.}~\bibnamefont{Hu}},
  \bibinfo{author}{\bibfnamefont{D.}~\bibnamefont{Weber}},
  \bibnamefont{et~al.}, \bibinfo{journal}{Nat. Commun.}
  \textbf{\bibinfo{volume}{11}}, \bibinfo{pages}{3879} (\bibinfo{year}{2020}).

\bibitem[{\citenamefont{Liu et~al.}(2019)\citenamefont{Liu, Wu, Tong, Li, Tao,
  and Petrovic}}]{liu2019}
\bibinfo{author}{\bibfnamefont{Y.}~\bibnamefont{Liu}},
  \bibinfo{author}{\bibfnamefont{L.}~\bibnamefont{Wu}},
  \bibinfo{author}{\bibfnamefont{X.}~\bibnamefont{Tong}},
  \bibinfo{author}{\bibfnamefont{J.}~\bibnamefont{Li}},
  \bibinfo{author}{\bibfnamefont{J.}~\bibnamefont{Tao}}, \bibnamefont{and}
  \bibinfo{author}{\bibfnamefont{C.}~\bibnamefont{Petrovic}},
  \bibinfo{journal}{Sci. Rep.} \textbf{\bibinfo{volume}{9}},
  \bibinfo{pages}{13599} (\bibinfo{year}{2019}).

\bibitem[{\citenamefont{Zorzi and Perottoni}(2021)}]{janete2021}
\bibinfo{author}{\bibfnamefont{J.~E.} \bibnamefont{Zorzi}} \bibnamefont{and}
  \bibinfo{author}{\bibfnamefont{C.}~\bibnamefont{Perottoni}},
  \bibinfo{journal}{Com. Mat. Sci.} \textbf{\bibinfo{volume}{199}},
  \bibinfo{pages}{110719} (\bibinfo{year}{2021}).

\bibitem[{\citenamefont{Spurgeon et~al.}(2020)\citenamefont{Spurgeon,
  Kozlowski, Susner, Turgut, and Boeckl}}]{spurgeon2020}
\bibinfo{author}{\bibfnamefont{K.}~\bibnamefont{Spurgeon}},
  \bibinfo{author}{\bibfnamefont{G.}~\bibnamefont{Kozlowski}},
  \bibinfo{author}{\bibfnamefont{M.~A.} \bibnamefont{Susner}},
  \bibinfo{author}{\bibfnamefont{Z.}~\bibnamefont{Turgut}}, \bibnamefont{and}
  \bibinfo{author}{\bibfnamefont{J.}~\bibnamefont{Boeckl}},
  \bibinfo{journal}{SCIREA Journal of Electric Engineering}
  \textbf{\bibinfo{volume}{5}}, \bibinfo{pages}{141} (\bibinfo{year}{2020}).

\bibitem[{\citenamefont{Lorenz et~al.}(2007)\citenamefont{Lorenz, Stark, Heyer,
  Hollmann, Vasiliev, Oosawa, and Tanaka}}]{lorenz2007}
\bibinfo{author}{\bibfnamefont{T.}~\bibnamefont{Lorenz}},
  \bibinfo{author}{\bibfnamefont{S.}~\bibnamefont{Stark}},
  \bibinfo{author}{\bibfnamefont{O.}~\bibnamefont{Heyer}},
  \bibinfo{author}{\bibfnamefont{N.}~\bibnamefont{Hollmann}},
  \bibinfo{author}{\bibfnamefont{A.}~\bibnamefont{Vasiliev}},
  \bibinfo{author}{\bibfnamefont{A.}~\bibnamefont{Oosawa}}, \bibnamefont{and}
  \bibinfo{author}{\bibfnamefont{H.}~\bibnamefont{Tanaka}},
  \bibinfo{journal}{J. Magn. Magn. Mat.} \textbf{\bibinfo{volume}{316}},
  \bibinfo{pages}{291} (\bibinfo{year}{2007}).

\bibitem[{\citenamefont{Heyer et~al.}(2011)\citenamefont{Heyer, Link, Wandner,
  Ruschewitz, and Lorenz}}]{heyer2011}
\bibinfo{author}{\bibfnamefont{O.}~\bibnamefont{Heyer}},
  \bibinfo{author}{\bibfnamefont{P.}~\bibnamefont{Link}},
  \bibinfo{author}{\bibfnamefont{D.}~\bibnamefont{Wandner}},
  \bibinfo{author}{\bibfnamefont{U.}~\bibnamefont{Ruschewitz}},
  \bibnamefont{and} \bibinfo{author}{\bibfnamefont{T.}~\bibnamefont{Lorenz}},
  \bibinfo{journal}{New J. Phys.} \textbf{\bibinfo{volume}{13}},
  \bibinfo{pages}{113041} (\bibinfo{year}{2011}).

\bibitem[{\citenamefont{Klingeler et~al.}(2006)\citenamefont{Klingeler, Geck,
  Arumugam, Tristan, Reutler, Büchner, Pinsard-Gaudart, and
  Revcolevschi}}]{klingeler2006}
\bibinfo{author}{\bibfnamefont{R.}~\bibnamefont{Klingeler}},
  \bibinfo{author}{\bibfnamefont{J.}~\bibnamefont{Geck}},
  \bibinfo{author}{\bibfnamefont{S.}~\bibnamefont{Arumugam}},
  \bibinfo{author}{\bibfnamefont{N.}~\bibnamefont{Tristan}},
  \bibinfo{author}{\bibfnamefont{P.}~\bibnamefont{Reutler}},
  \bibinfo{author}{\bibfnamefont{B.}~\bibnamefont{Büchner}},
  \bibinfo{author}{\bibfnamefont{L.}~\bibnamefont{Pinsard-Gaudart}},
  \bibnamefont{and}
  \bibinfo{author}{\bibfnamefont{A.}~\bibnamefont{Revcolevschi}},
  \bibinfo{journal}{Phys. Rev. B} \textbf{\bibinfo{volume}{73}},
  \bibinfo{pages}{214432} (\bibinfo{year}{2006}).

\bibitem[{\citenamefont{Gegenwart}(2016)}]{gegenwart2016}
\bibinfo{author}{\bibfnamefont{P.}~\bibnamefont{Gegenwart}},
  \bibinfo{journal}{Rep. Prog. Phys.} \textbf{\bibinfo{volume}{79}},
  \bibinfo{pages}{114502} (\bibinfo{year}{2016}).

\bibitem[{\citenamefont{Hoffmann et~al.}(2021)\citenamefont{Hoffmann, Dey,
  Werner, Bag, Kaiser, Wadepohl, Skourski, Abdel-Hafiez, Singh, and
  Klingeler}}]{hoffmann2021}
\bibinfo{author}{\bibfnamefont{M.}~\bibnamefont{Hoffmann}},
  \bibinfo{author}{\bibfnamefont{K.}~\bibnamefont{Dey}},
  \bibinfo{author}{\bibfnamefont{J.}~\bibnamefont{Werner}},
  \bibinfo{author}{\bibfnamefont{R.}~\bibnamefont{Bag}},
  \bibinfo{author}{\bibfnamefont{J.}~\bibnamefont{Kaiser}},
  \bibinfo{author}{\bibfnamefont{H.}~\bibnamefont{Wadepohl}},
  \bibinfo{author}{\bibfnamefont{Y.}~\bibnamefont{Skourski}},
  \bibinfo{author}{\bibfnamefont{M.}~\bibnamefont{Abdel-Hafiez}},
  \bibinfo{author}{\bibfnamefont{S.}~\bibnamefont{Singh}}, \bibnamefont{and}
  \bibinfo{author}{\bibfnamefont{R.}~\bibnamefont{Klingeler}},
  \bibinfo{journal}{Phys. Rev. B} \textbf{\bibinfo{volume}{104}},
  \bibinfo{pages}{014429} (\bibinfo{year}{2021}).

\bibitem[{\citenamefont{Küchler}(2005)}]{kuechlerDiss}
\bibinfo{author}{\bibfnamefont{R.}~\bibnamefont{Küchler}}, Ph.D. thesis,
  \bibinfo{school}{Technische Universität Dresden} (\bibinfo{year}{2005}).

\bibitem[{\citenamefont{Meingast et~al.}(1990)\citenamefont{Meingast, Blank,
  Bürkle, Obst, Wolf, Wühl, Selvamanickam, and Salama}}]{meingast1990}
\bibinfo{author}{\bibfnamefont{C.}~\bibnamefont{Meingast}},
  \bibinfo{author}{\bibfnamefont{B.}~\bibnamefont{Blank}},
  \bibinfo{author}{\bibfnamefont{H.}~\bibnamefont{Bürkle}},
  \bibinfo{author}{\bibfnamefont{B.}~\bibnamefont{Obst}},
  \bibinfo{author}{\bibfnamefont{T.}~\bibnamefont{Wolf}},
  \bibinfo{author}{\bibfnamefont{H.}~\bibnamefont{Wühl}},
  \bibinfo{author}{\bibfnamefont{V.}~\bibnamefont{Selvamanickam}},
  \bibnamefont{and} \bibinfo{author}{\bibfnamefont{K.}~\bibnamefont{Salama}},
  \bibinfo{journal}{Phys. Rev. B} \textbf{\bibinfo{volume}{41}},
  \bibinfo{pages}{11299} (\bibinfo{year}{1990}).

\bibitem[{\citenamefont{Majumder et~al.}(2018)\citenamefont{Majumder, Manna,
  Simutis, Orain, Dey, Freund, Jesche, Khasanov, Biswas, Bykova
  et~al.}}]{majumder2018}
\bibinfo{author}{\bibfnamefont{M.}~\bibnamefont{Majumder}},
  \bibinfo{author}{\bibfnamefont{R.~S.} \bibnamefont{Manna}},
  \bibinfo{author}{\bibfnamefont{G.}~\bibnamefont{Simutis}},
  \bibinfo{author}{\bibfnamefont{J.~C.} \bibnamefont{Orain}},
  \bibinfo{author}{\bibfnamefont{T.}~\bibnamefont{Dey}},
  \bibinfo{author}{\bibfnamefont{F.}~\bibnamefont{Freund}},
  \bibinfo{author}{\bibfnamefont{A.}~\bibnamefont{Jesche}},
  \bibinfo{author}{\bibfnamefont{R.}~\bibnamefont{Khasanov}},
  \bibinfo{author}{\bibfnamefont{P.~K.} \bibnamefont{Biswas}},
  \bibinfo{author}{\bibfnamefont{E.}~\bibnamefont{Bykova}},
  \bibnamefont{et~al.}, \bibinfo{journal}{Phys. Rev. Lett.}
  \textbf{\bibinfo{volume}{120}}, \bibinfo{pages}{237202}
  (\bibinfo{year}{2018}).

\bibitem[{\citenamefont{Ghosh et~al.}()\citenamefont{Ghosh, Singh, Mu,
  Kvashnin, Haider, Jonak, Chareev, Aramaki, Medvedev, Klingeler
  et~al.}}]{ghosh2021arxiv}
\bibinfo{author}{\bibfnamefont{A.}~\bibnamefont{Ghosh}},
  \bibinfo{author}{\bibfnamefont{D.}~\bibnamefont{Singh}},
  \bibinfo{author}{\bibfnamefont{Q.}~\bibnamefont{Mu}},
  \bibinfo{author}{\bibfnamefont{Y.}~\bibnamefont{Kvashnin}},
  \bibinfo{author}{\bibfnamefont{G.}~\bibnamefont{Haider}},
  \bibinfo{author}{\bibfnamefont{M.}~\bibnamefont{Jonak}},
  \bibinfo{author}{\bibfnamefont{D.}~\bibnamefont{Chareev}},
  \bibinfo{author}{\bibfnamefont{T.}~\bibnamefont{Aramaki}},
  \bibinfo{author}{\bibfnamefont{S.~A.} \bibnamefont{Medvedev}},
  \bibinfo{author}{\bibfnamefont{R.}~\bibnamefont{Klingeler}},
  \bibnamefont{et~al.}, \bibinfo{note}{arXiv:2108.00173}.

\bibitem[{\citenamefont{Mondal et~al.}(2019)\citenamefont{Mondal, Kannan, Das,
  Govindaraj, Singha, Satpati, Arumugam, and Mandal}}]{mondal2019}
\bibinfo{author}{\bibfnamefont{S.}~\bibnamefont{Mondal}},
  \bibinfo{author}{\bibfnamefont{M.}~\bibnamefont{Kannan}},
  \bibinfo{author}{\bibfnamefont{M.}~\bibnamefont{Das}},
  \bibinfo{author}{\bibfnamefont{L.}~\bibnamefont{Govindaraj}},
  \bibinfo{author}{\bibfnamefont{R.}~\bibnamefont{Singha}},
  \bibinfo{author}{\bibfnamefont{B.}~\bibnamefont{Satpati}},
  \bibinfo{author}{\bibfnamefont{S.}~\bibnamefont{Arumugam}}, \bibnamefont{and}
  \bibinfo{author}{\bibfnamefont{P.}~\bibnamefont{Mandal}},
  \bibinfo{journal}{Phys. Rev. B} \textbf{\bibinfo{volume}{99}},
  \bibinfo{pages}{180407} (\bibinfo{year}{2019}).

\bibitem[{\citenamefont{He et~al.}(2018)\citenamefont{He, Wang, Wang, Hardy,
  Wolf, Adelmann, Brückel, Su, and Meingast}}]{he2018}
\bibinfo{author}{\bibfnamefont{M.}~\bibnamefont{He}},
  \bibinfo{author}{\bibfnamefont{X.}~\bibnamefont{Wang}},
  \bibinfo{author}{\bibfnamefont{L.}~\bibnamefont{Wang}},
  \bibinfo{author}{\bibfnamefont{F.}~\bibnamefont{Hardy}},
  \bibinfo{author}{\bibfnamefont{T.}~\bibnamefont{Wolf}},
  \bibinfo{author}{\bibfnamefont{P.}~\bibnamefont{Adelmann}},
  \bibinfo{author}{\bibfnamefont{T.}~\bibnamefont{Brückel}},
  \bibinfo{author}{\bibfnamefont{Y.}~\bibnamefont{Su}}, \bibnamefont{and}
  \bibinfo{author}{\bibfnamefont{C.}~\bibnamefont{Meingast}},
  \bibinfo{journal}{J. Phys.: Condens. Matter} \textbf{\bibinfo{volume}{30}},
  \bibinfo{pages}{385702} (\bibinfo{year}{2018}).

\bibitem[{\citenamefont{Li et~al.}(2019)\citenamefont{Li, Jiang, Sivadas, Wang,
  Xu, Weber, Goldberger, Watanabe, Taniguchi, Fennie et~al.}}]{li2019}
\bibinfo{author}{\bibfnamefont{T.}~\bibnamefont{Li}},
  \bibinfo{author}{\bibfnamefont{S.}~\bibnamefont{Jiang}},
  \bibinfo{author}{\bibfnamefont{N.}~\bibnamefont{Sivadas}},
  \bibinfo{author}{\bibfnamefont{Z.}~\bibnamefont{Wang}},
  \bibinfo{author}{\bibfnamefont{Y.}~\bibnamefont{Xu}},
  \bibinfo{author}{\bibfnamefont{D.}~\bibnamefont{Weber}},
  \bibinfo{author}{\bibfnamefont{J.}~\bibnamefont{Goldberger}},
  \bibinfo{author}{\bibfnamefont{K.}~\bibnamefont{Watanabe}},
  \bibinfo{author}{\bibfnamefont{T.}~\bibnamefont{Taniguchi}},
  \bibinfo{author}{\bibfnamefont{C.~J.} \bibnamefont{Fennie}},
  \bibnamefont{et~al.}, \bibinfo{journal}{Nat. Mater.}
  \textbf{\bibinfo{volume}{18}}, \bibinfo{pages}{1303} (\bibinfo{year}{2019}).

\bibitem[{\citenamefont{León et~al.}(2020)\citenamefont{León, González,
  Mejía-López, de~Lima, and Morell}}]{leon2020}
\bibinfo{author}{\bibfnamefont{A.~M.} \bibnamefont{León}},
  \bibinfo{author}{\bibfnamefont{J.~W.} \bibnamefont{González}},
  \bibinfo{author}{\bibfnamefont{J.}~\bibnamefont{Mejía-López}},
  \bibinfo{author}{\bibfnamefont{F.~C.} \bibnamefont{de~Lima}},
  \bibnamefont{and} \bibinfo{author}{\bibfnamefont{E.~S.}
  \bibnamefont{Morell}}, \bibinfo{journal}{2D Mater.}
  \textbf{\bibinfo{volume}{7}}, \bibinfo{pages}{035008} (\bibinfo{year}{2020}).

\bibitem[{\citenamefont{Dupont et~al.}(2021)\citenamefont{Dupont, Kvashnin,
  Shiranzaei, Fransson, Laflorencie, and Kantian}}]{dupont2020}
\bibinfo{author}{\bibfnamefont{M.}~\bibnamefont{Dupont}},
  \bibinfo{author}{\bibfnamefont{Y.~O.} \bibnamefont{Kvashnin}},
  \bibinfo{author}{\bibfnamefont{M.}~\bibnamefont{Shiranzaei}},
  \bibinfo{author}{\bibfnamefont{J.}~\bibnamefont{Fransson}},
  \bibinfo{author}{\bibfnamefont{N.}~\bibnamefont{Laflorencie}},
  \bibnamefont{and} \bibinfo{author}{\bibfnamefont{A.}~\bibnamefont{Kantian}},
  \bibinfo{journal}{Phys. Rev. Lett.} \textbf{\bibinfo{volume}{127}},
  \bibinfo{pages}{037204} (\bibinfo{year}{2021}).

\bibitem[{\citenamefont{Anderson}(1959)}]{anderson1959}
\bibinfo{author}{\bibfnamefont{P.~W.} \bibnamefont{Anderson}},
  \bibinfo{journal}{Phys. Rev.} \textbf{\bibinfo{volume}{115}},
  \bibinfo{pages}{2} (\bibinfo{year}{1959}).

\bibitem[{\citenamefont{Kanamori}(1959)}]{kanamori1959}
\bibinfo{author}{\bibfnamefont{J.}~\bibnamefont{Kanamori}},
  \bibinfo{journal}{J. Phys. Chem. Solids} \textbf{\bibinfo{volume}{10}},
  \bibinfo{pages}{87} (\bibinfo{year}{1959}).

\bibitem[{\citenamefont{Goodenough}(1958)}]{goodenough1958}
\bibinfo{author}{\bibfnamefont{J.~B.} \bibnamefont{Goodenough}},
  \bibinfo{journal}{J. Phys. Chem. Solids} \textbf{\bibinfo{volume}{6}},
  \bibinfo{pages}{287} (\bibinfo{year}{1958}).

\bibitem[{\citenamefont{Kashin et~al.}(2020)\citenamefont{Kashin, Mazurenko,
  Katsnelson, and Rudenko}}]{kashin2020}
\bibinfo{author}{\bibfnamefont{I.~V.} \bibnamefont{Kashin}},
  \bibinfo{author}{\bibfnamefont{V.~V.} \bibnamefont{Mazurenko}},
  \bibinfo{author}{\bibfnamefont{M.~I.} \bibnamefont{Katsnelson}},
  \bibnamefont{and} \bibinfo{author}{\bibfnamefont{A.~N.}
  \bibnamefont{Rudenko}}, \bibinfo{journal}{2D Mater.}
  \textbf{\bibinfo{volume}{7}}, \bibinfo{pages}{025036} (\bibinfo{year}{2020}).

\bibitem[{\citenamefont{Song et~al.}(2019)\citenamefont{Song, Fei, Yankowitz,
  Lin, Jiang, Hwangbo, Zhang, Sun, Taniguchi, Watanabe et~al.}}]{song2019}
\bibinfo{author}{\bibfnamefont{T.}~\bibnamefont{Song}},
  \bibinfo{author}{\bibfnamefont{Z.}~\bibnamefont{Fei}},
  \bibinfo{author}{\bibfnamefont{M.}~\bibnamefont{Yankowitz}},
  \bibinfo{author}{\bibfnamefont{Z.}~\bibnamefont{Lin}},
  \bibinfo{author}{\bibfnamefont{Q.}~\bibnamefont{Jiang}},
  \bibinfo{author}{\bibfnamefont{K.}~\bibnamefont{Hwangbo}},
  \bibinfo{author}{\bibfnamefont{Q.}~\bibnamefont{Zhang}},
  \bibinfo{author}{\bibfnamefont{B.}~\bibnamefont{Sun}},
  \bibinfo{author}{\bibfnamefont{T.}~\bibnamefont{Taniguchi}},
  \bibinfo{author}{\bibfnamefont{K.}~\bibnamefont{Watanabe}},
  \bibnamefont{et~al.}, \bibinfo{journal}{Nat. Mater.}
  \textbf{\bibinfo{volume}{18}}, \bibinfo{pages}{1298–1302}
  (\bibinfo{year}{2019}).

\end{thebibliography}

\clearpage
\newpage

\beginsupplement
\onecolumngrid

\section*{\large{Supplemental material: }}

\begin{figure}[h]
    \centering
    \includegraphics[width = 0.6\textwidth]{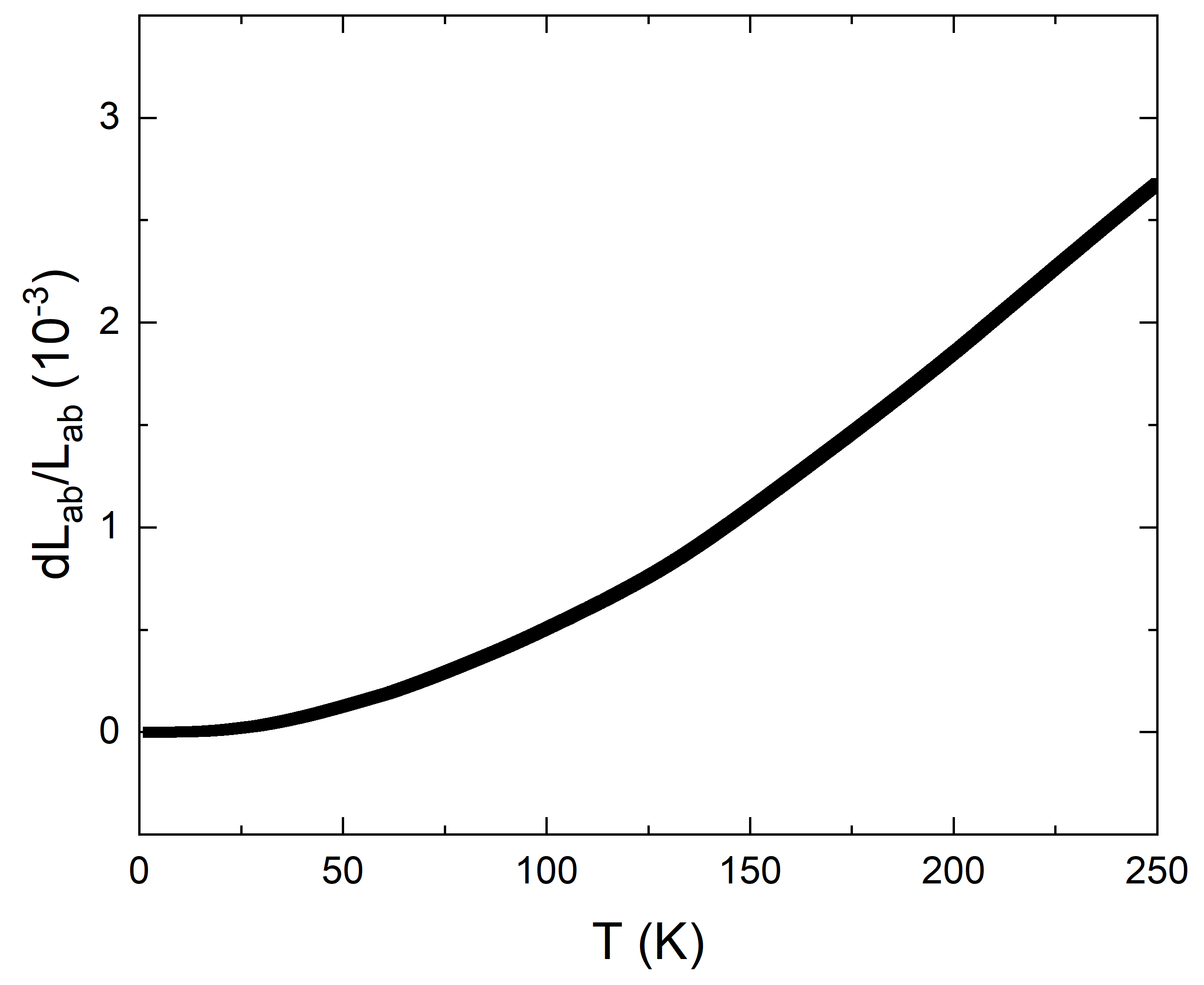}
    \caption{In-plane relative length changes in \cri{} from $2\,\mathrm{K}$ to $250\,\mathrm{K}$. The data show no discontinuity at the structural phase transition at $T_\mathrm{S} \approx 220\,\mathrm{K}$.}
    \label{fig:dLL_ab}
\end{figure}

\end{document}